\documentclass[a4paper]{article}
\usepackage{fullpage}
\usepackage{graphicx}
\usepackage{float}
\usepackage{rotating}
\usepackage[font=small]{caption}
\usepackage{amsfonts}
\usepackage{amssymb,amsmath}

\usepackage{authblk}

\usepackage[font={small,it}]{caption}
\usepackage{wrapfig}
\usepackage{setspace}
\usepackage{natbib}
\usepackage{amsthm}
\newtheorem{Definition}{Definition}

\usepackage{mathtools}

\usepackage{marvosym}
\usepackage{subcaption}

\DeclareMathOperator{\diag}{diag}

\begin{document}
\title{}
\date{}
\author{}
{\Large\bf Network Inference and Community Detection, Based on Covariance Matrices, Correlations and Test Statistics from Arbitrary Distributions}
\vspace{1ex}
\\
Thomas E. Bartlett$^{1\ast}$
\vspace{1ex}
\\
1. Department of Statistical Science, University College London, London WC1E 7HB, United Kingdom.
\\
$\ast$ E-mail: thomas.bartlett.10@ucl.ac.uk

\begin{abstract}
In this paper we propose methodology for inference of binary-valued adjacency matrices from various measures of the strength of association between pairs of network nodes, or more generally pairs of variables. This strength of association can be quantified by sample covariance and correlation matrices, and more generally by test-statistics and hypothesis test \textit{p}-values from arbitrary distributions. Community detection methods such as block modelling typically require binary-valued adjacency matrices as a starting point. Hence, a main motivation for the methodology we propose is to obtain binary-valued adjacency matrices from such pairwise measures of strength of association between variables. The proposed methodology is applicable to large high-dimensional data-sets and is based on computationally efficient algorithms. We illustrate its utility in a range of contexts and data-sets.
\end{abstract}

\section{Introduction}
Networks and other non-Euclidean relational datasets have become important applications in modern statistics. Key considerations include balancing statistical fidelity with computational tractability. Much effort has gone into developing parametric models for networks which take account of such considerations, typically by specifying both node-specific effects such as degree, and grouped-node effects such as community structure \citep{holland1983stochastic,bickel2009nonparametric,rohe2011spectral,qin2013regularized,wilson2013testing}. One of the most widely studied of these models is the stochastic blockmodel in which (under the assortative assumption) there is a greater probability of observing an edge (or interaction) between a pair of nodes (or entities) if they are in the same block, or community. Practical approaches to finding communities in social and biological networks have been studied for many years \citep{girvan2002community}, and real life examples of this problem include identifying groups of friends in social networks, and identifying functional subnetwork modules in biological networks. In the biological setting, considering groups of genes defined together as subgraphs can lead to increases in statistical power, aiding discovery of biological phenomena \citep{jacob2012more,li2010variable,peng2010regularized}. 

There are important differences between community detection and clustering. A community within a network typically refers to a grouping of entities with a strong tendency for direct interaction within the group, such as a friendship group in a social network. On the other hand, a cluster typically refers to a group of variables which are highly correlated, but these variables do not necessarily represent entities which interact directly. However, practical application of community detection and clustering methodologies often yield similar results. The stochastic blockmodel is an efficient method to detect communities in networks, and more generally it can be used to cluster together variables with correlated observations. However, most of the important theoretical understanding of the stochastic blockmodel has been developed under the assumption of a binary-valued relationship between the network nodes \citep{holland1983stochastic,bickel2009nonparametric,rohe2011spectral,qin2013regularized,wilson2013testing,olhede2014net}. This relationship corresponds to the presence and absence of network edges between these nodes, and is typically represented ones and zeros (respectively) in an adjacency matrix. If such theoretical understanding is to be relevant to the use of community detection / the stochastic blockmodel as a means of clustering, the data to be clustered must first be transformed into this binary-valued format.

The methodology that we propose in this paper allows a binary-valued adjacency matrix to be estimated based on association matrices composed of sample covariances, or correlations, or test statistics from arbitrary known or unknown distributions. This binary-valued adjacency matrix is then an ideal summary of the relational data-set on which to carry out community detection. Hence, the main motivation of this paper is to propose methodology to allow continuous-valued statistics which measure the strength of association between pairs of variables to be transformed into a binary-valued adjacency matrix format, for use in community detection. In this format, ones and zeros can be considered to represent variables which are and are not correlated, respectively.

If a binary-valued adjacency matrix is used to define pairs of variables which are correlated, and other pairs of variables which are not correlated, then the zero entries in this matrix define pairs of variables which are independent. This relates closely to the `probabilistic graphical model' \citep{koller2009probabilistic} paradigm, in which a joint probability distribution over a large number of variables is made tractable by taking advantage of independencies between pairs of variables as specified by the graphical model. These ideas are also closely related to thresholding a covariance matrix to a sparse representation \citep{bickel2008covariance,rothman2009generalized,bien2011sparse}, where again zeros in the sparse representation imply independent pairs of variables. Sparse multivariate methods such as the lasso \citep{tibshirani1996regression} are also popular for obtaining sparse representations via linear modelling, and can be extended to networks data via the graphical lasso \cite{friedman2008sparse}. However the methodology proposed in this paper offers two main advantages over the lasso in this context. Firstly, the computational implementation is via a closed-form expression and therefore it is much quicker than the iterative procedures required by the lasso. Secondly, the mixture-modelling strategy we employ is precisely specified for the problem we consider here, unlike the lasso.

This paper is organised as follows. In section \ref{adjMatEst} we define notation and present the methodology and practical details for its usage and implementation. Then in section \ref{examples}, we present examples to illustrate the performance of this methodology, including a simulation study and several real data-sets from different contexts.

\section{Proposed methodology}\label{adjMatEst}
We start this section by specifying the model which we will use to estimate the adjacency matrix $\mathbf{A}$.
\begin{Definition}\label{mainDefStatement}
For $m\in\mathbb{N}^+$ define the set of network nodes $\{1,...,m\}$, and for each node $i$ define a corresponding variable $x_i$. Let ${z}_{ij}$ represent an observed measure of association/dependence between variables $x_i$ and $x_j$, where: 
\begin{equation*}
{z}_{ij}\sim\mathcal{N}\left(\mu_{ij},\sigma^2\right).
\end{equation*}
Let $\mathbf{A}\in\left\{0,1\right\}^{m\times m}$ be an adjacency matrix, the elements of which satisfy:
\begin{align*}
   A_{ij}=&\begin{cases}
    0, & \text{if there is no edge between nodes }i\text{ and }j\text{, implying}\\
    	& \text{that the variables }x_i\text{ and }x_j\text{ are independent},\\
	& \\
    1, & \text{if there is an edge between nodes }i\text{ and }j\text{, implying}\\
    	& \text{that the variables }x_i\text{ and }x_j\text{ are not independent},
\end{cases}
\end{align*}
and let $w=p\left(A_{ij}=1\right)$. Then, the observed measures of association ${z}_{ij}$ may be modelled using the mixture distribution:
\begin{equation}
{z}_{ij}\sim (1-w)\cdot\mathcal{N}\left(0,\sigma^2\right)+w\cdot\mathcal{N}\left(\mu_{ij},\sigma^2\right).\label{mainDef}
\end{equation}
\end{Definition}
\noindent 
In section \ref{covCorSubSect} we describe how to calculate the observed measures of association/dependence ${z}_{ij}$ from sample covariance/correlation matrices. Then, in section \ref{pValAppSubSect}, we describe the equivalent calculations based on test statistics from arbitrary or unknown distributions. Next, in section \ref{modelFitSect} we describe how the model of definition \ref{mainDefStatement} can be fitted, and how the adjacency matrix $\hat{\mathbf{A}}$ can be estimated from the fitted model. Then in section \ref{comDetSubSect}, we discuss community detection based on $\hat{\mathbf{A}}$. 

\subsection{Applying the model to a covariance/correlation matrix}\label{covCorSubSect}
We can estimate an adjacency matrix from a sample covariance or correlation matrix by fitting the model of definition \ref{mainDefStatement} by starting with the following procedure. Equation \ref{covMat} defines the sample covariance matrix $\hat{\boldsymbol{\Sigma}}$ for the $m$ variables represented by the vector $\mathbf{x}$, $x_1,...,x_m$, for samples $\mathbf{x}(k)$, $k=1,...,n$:
\begin{equation}
\hat{\boldsymbol{\Sigma}}=\frac{1}{n}{\sum_{k=1}^{n}\left(\mathbf{x}(k)-\bar{\mathbf{x}}\right)\left(\mathbf{x}(k)-\bar{\mathbf{x}}\right)^T}\label{covMat},\qquad\text{where}\quad\bar{\mathbf{x}}=\frac{1}{n}\sum_{k=1}^{n}\mathbf{x}(k).
\end{equation}
By dividing each row and each column of $\hat{\boldsymbol{\Sigma}}$ by the square roots of the corresponding elements of the leading diagonal, we obtain the sample correlation matrix $\hat{\boldsymbol{r}}$:
\begin{equation*}
\hat{\boldsymbol{r}}=\left(\diag(\hat{\boldsymbol{\Sigma}})\right)^{-1/2}\hat{\boldsymbol{\Sigma}}\left(\diag(\hat{\boldsymbol{\Sigma}})\right)^{-1/2}.
\end{equation*}
The $(i,j)^\text{th}$ element of $\hat{\boldsymbol{r}}$, i.e. $\hat{r}_{ij}$, is the Pearson correlation coefficient between variables $x_i$ and $x_j$. If $x_i$ and $x_j$ are jointly normally distributed, and the $\left\{x_i(k),x_j(k)\right\}$, $k=1,...,n$ samples are independent, the Fisher transform \citep{fisher1915frequency} converts $\hat{r}_{ij}$ to the approximately normally distributed variable ${z}_{ij}$:
\begin{equation}
{z}_{ij}=\frac{1}{2}\ln\left(\frac{1+\hat{r}_{ij}}{1-\hat{r}_{ij}}\right),\label{fisherTrans}
\end{equation}
where 
\begin{equation*}
{z}_{ij}\overset{approx}{\sim}\mathcal{N}\left(\frac{1}{2}\ln\left(\frac{1+r_{ij}}{1-r_{ij}}\right),\frac{1}{\nu-3}\right),
\end{equation*}
where $r_{ij}$ is the true correlation coefficient between variables $x_i$ and $x_j$, and $\nu$ is the degrees of freedom. Hence, we can model the Fisher-transformed sample correlation coefficients ${z}_{ij}$ with the mixture model of equation \ref{mainDef}, also with:
\begin{equation}
\quad\mu_{ij}=\frac{1}{2}\ln\left(\frac{1+r_{ij}}{1-r_{ij}}\right)\quad\text{and}\quad\sigma^2=\frac{1}{\nu-3}\label{fisherMix}.
\end{equation}

\subsection{Applying the model to test statistics from arbitrary distributions}\label{pValAppSubSect}
We can also estimate an adjacency matrix by fitting the model of definition \ref{mainDefStatement} when the association between variables $x_i$ and $x_j$ is assessed by a test-statistic from an arbitrary distribution expressed as a hypothesis-test \textit{p}-value. Such a \textit{p}-values may result from test-statistics from any known distribution, or may even be derived from an unknown distribution, for example by Monte-Carlo simulation. We can represent these \textit{p}-values in the matrix ${\mathbf{P}}$, where ${p}_{ij}$ is the estimated probability of observing the association test-statistic for the pair of variables $x_i$ and $x_j$ under the null hypothesis $H_0$ that there is no association between $x_i$ and $x_j$ (i.e. they are independent). Assuming these \textit{p}-values arose from upper-tailed tests, we can apply the inverse-normal transformation as follows:
\begin{equation}
{z}_{ij}=\Phi^{-1}\left(1-{p}_{ij}\right),\label{invNormTrans}
\end{equation}
with an equivalent expression available for lower-tailed tests. Applying this transformation is equivalent to applying quantile normalisation, mapping the null distribution of $p_{ij}$ onto the standard normal $\mathcal{N}\left(0,1\right)$ distribution. Hence, after applying this transformation we can again fit the mixture model of definition \ref{mainDefStatement}, and use this model fit to infer the estimated adjacency matrix $\hat{\mathbf{A}}$.

\subsection{Model fitting and adjacency matrix inference}\label{modelFitSect}
We propose fitting the model of definition \ref{mainDefStatement} with an empirical Bayes procedure used previously for thresholding \citep{johnstone2004needles}. This method is based on a mixture prior over $\mu_{ij}$, with a Laplace density for the non-zero mean component.
\begin{Definition}\label{eBayesMixDefStatement}
With $\mu_{ij}$ and $w$ given by definition \ref{mainDefStatement}, let $\gamma\left(\cdot\right)$ represent the Laplace distribution probability density function with spread parameter $a$:
\begin{equation*}
\gamma\left(\mu_{ij}\right)=\frac{a}{2}\exp{\left(-a\left|\mu_{ij}\right|\right)}.
\end{equation*}
Then, the mixture prior over $\mu_{ij}$ is defined as:
\begin{equation*}
f_\text{prior}\left(\mu_{ij}\right)=\left(1-w_i\right)\delta\left(\mu_{ij}\right)+w_i\gamma\left(\mu_{ij}\right).
\end{equation*}
\end{Definition}
\noindent
Typically the Laplace spread parameter is taken as $a=0.5$. If the mixture components have Gaussian likelihoods $f_\mathcal{N}\left(\cdot\middle|\mu_{ij},\sigma^2\right)$ as in definition \ref{mainDefStatement}, it follows from definition \ref{eBayesMixDefStatement} that the posterior density over the observed measures of association ${z}_{ij}$ is:
\begin{equation*}
f_\text{posterior}\left(\mu_{ij}\middle|{z}_{ij}\right)=\frac{\left(1-w_i\right)\delta\left(\mu_{ij}\right)f_\mathcal{N}\left({z}_{ij}\middle|0,\sigma^2\right)+w_i\gamma\left(\mu_{ij}\right)f_\mathcal{N}\left({z}_{ij}\middle|\mu_{ij},\sigma^2\right)}{f_\text{marginal}\left({z}_{ij}\right)},
\end{equation*}
where the marginal density is:
\begin{equation}
f_\text{marginal}\left({z}_{ij}\right)=(1-w_i)f_\mathcal{N}\left({z}_{ij}\middle|0,\sigma^2\right)+w_ig\left({z}_{ij}\right),\label{margDist}
\end{equation}
where $g\left(\mu_{ij}\right)$ is the convolution of the Laplace density with the standard normal density. Comparing the expression for $f_\text{marginal}\left({z}_{ij}\right)$ in equation \ref{margDist} with equation \ref{mainDef}, we see that the normally-distributed non-zero mean mixture component in equation \ref{mainDef} is replaced with the convolution of this Laplace and normal densities in equation \ref{margDist}. If a Gaussian prior were used here instead of the Laplace prior, then the marginal density in equation \ref{margDist} would be exactly the same as equation \ref{mainDef}. However, as noted previously \citep{johnstone2004needles}, this empirical Bayes procedure requires a prior with tails that are exponential or heavier. Hence we use, as previously, the Laplace rather than a Gaussian prior. We note that this is a slight model mis-specification.

This procedure results in a separate model being fitted to each pair of variables $(x_i,x_j)$, based on the corresponding observed statistic ${z}_{ij}$. This methodology was originally developed to be applied to vector data (in the form of wavelet coefficients) \citep{johnstone2004needles}. Because the dependency structure of matrix data (such as covariance or correlation matrices) may be different to that of vector data, we apply the model fitting to each row of the association matrix, i.e. a vector, separately. As the association matrices under consideration are symmetric, this is equivalent to applying the method to both rows and columns of the matrix. We then take a conservative estimate, only inferring an edge in the network when there is agreement between the result of model fitting with respect to both rows and columns of the association matrix. Applying the methodology in this way results in a common weight $w_i$ being used for all models corresponding to each $x_i$. This estimate of $w_i$ is found as the value which maximises the marginal likelihood (equation \ref{margMLE}) of the observed statistics ${z}_{ij}$ over all the pairwise comparisons of $x_i$ with $x_j$, $j\neq i$. This allows the model for each pairwise comparison $(x_i,x_j)$ to `borrow strength' from all the other comparisons $(x_i,x_{j'})$, $j'\neq i$, $j'\neq j$: 
\begin{equation}
\hat{w}_i=\arg\max_w\sum_{j\neq i}\log\left\{(1-w)\phi\left({z}_{ij}\right)+wg\left({z}_{ij}\right)\right\}.\label{margMLE}
\end{equation}
For a particular $x_i$, if the ${z}_{ij}$ are mostly close to zero then $w_i$ will be set low, which means that fewer edges ($A_{ij}=1$) will be detected: this corresponds to $i$ being a low-degree node. If for a different $x_i$, the ${z}_{ij}$ are generally further from zero, then $\hat{w}_i$ will be set high, which corresponds to more edges being detected: this corresponds to $i$ being a high-degree node. Hence, setting $\hat{w}_i$ separately for each variable $x_i$ allows adaptation to a heterogenous degree distribution in $\mathbf{A}$.

As in the original use of this methodology \citep{johnstone2004needles}, we use the posterior median to calculate $\hat{\mu}_{ij}$. Based on this, we can estimate the corresponding adjacency matrix entry $A_{ij}$ as:
\begin{align}
\hat{A}_{ij}=&1\quad\text{if}\quad\left|\hat{\mu}_{ij}\right|>0,\label{eBayesMixMLEvec}\\
\hat{A}_{ij}=&0\quad\text{otherwise}.\nonumber
\end{align}
We make the conservative estimate of $A_{ij}$ discussed above as follows:
\begin{align}
\hat{A}_{ij}=&1\quad\text{if}\quad\left|\hat{\mu}_{ij}\right|>0\quad\text{and}\quad\left|\hat{\mu}_{ji}\right|>0,\label{eBayesMixMLE}\\
\hat{A}_{ij}=&0\quad\text{otherwise}.\nonumber
\end{align}
We note that requiring agreement between $\left|\hat{\mu}_{ij}\right|>0$ and $\left|\hat{\mu}_{ji}\right|>0$ is likely to result in decreased sensitivity: this point is discussed further in section \ref{simStudSect} the context of the simulation study. The spread parameter $a$ in the Laplace prior is set as standard as $a=0.5$ \citep{johnstone2004needles}. However, for additional model flexibility where needed, $a$ can also be estimated by marginal maximum likelihood, in which case we estimate $a_i$ separately for each variable $x_i$, simultaneously with $w_i$. 

\subsection{Community detection}\label{comDetSubSect}
Having inferred $\hat{\mathbf{A}}$, community detection \citep{girvan2002community} may then proceed by fitting the degree-corrected stochastic blockmodel \citep{holland1983stochastic,bickel2009nonparametric,rohe2011spectral,qin2013regularized} directly to $\hat{\mathbf{A}}$. However, to fit the degree-corrected stochastic blockmodel the number of communities in the model, $T$, must first be specified; this number can be estimated by the `network histogram' method \citep{olhede2014net}. Using this estimate of the number of communities, we infer the set of communities $\hat{C}$ based on $\hat{\mathbf{A}}$, such that a community $\hat{c}_t\in\hat{C}$, $t\in\left\{1,...,T\right\}$, is a group of variables $x_i$, $i\in\hat{c}_t$. Such a community $\hat{c}_t$ would correspond to an unexpectedly large number of non-zero entries $|\hat{\Sigma}_{ij}|>0$ of the sample covariance matrix $\hat{\boldsymbol{\Sigma}}$ for pairs of variables $x_i$ and $x_j$ where $i\in\hat{c}_t$ and $j\in\hat{c}_t$. Alternatively, the community $\hat{c}_t$ would correspond to an unexpectedly large number of significant \textit{p}-values ${p}_{ij}$ in the matrix ${\mathbf{P}}$ for pairs of variables $x_i$ and $x_j$ again with $i\in\hat{c}_t$ and $j\in\hat{c}_t$.

\section{Examples}\label{examples}
In this section, we present the results of applying the methodology proposed in section \ref{adjMatEst} to simulated data, and to publicly available data-sets relevant to genomics and consumer-product reviews. For each data-set, we carry out network inference as described in sections \ref{covCorSubSect} - \ref{modelFitSect} resulting in a binary-valued adjacency matrix. To each such adjacency matrix, we fit the degree-corrected stochastic blockmodel, by regularised spectral clustering \citep{holland1983stochastic,bickel2009nonparametric,rohe2011spectral,qin2013regularized}. Spectral clustering is in general computationally intensive, as it requires the singular value decomposition (SVD) of a large matrix. However, the network inference described in sections \ref{covCorSubSect} - \ref{modelFitSect} provides us with a sparse binary-valued adjacency matrix, and efficient computational methods exist to find the top few components in the SVD of large sparse matrices \citep{sorensen1992implicit,lehoucq1996deflation}. Hence, as we only require as many SVD components as the number of communities or clusters we are trying to find (which tends to be two or more orders of magnitude smaller than the dimension of the adjacency matrix, $m$), these efficient computational methods can be used here. Relevant software implementations of these methods are included in \textit{Matlab} and \textit{R}, meaning that this methodology is practical for large data-sets, and is quick to implement for many end-users.

\subsection{Simulation study}\label{simStudSect}
We first carried out a simulation study, to assess the effectiveness of our network inference methodology in the context of generated networks with known community structure. A generative model for exchangeable random networks with heterogenous degrees is the logistic-linear model \citep{perry2012null}. We use a version of that model here with community structure added, defined as:
\begin{equation*}
\text{Logit}\left(p_{ij}\right)=\alpha_i+\alpha_j+\theta_{ij}
\end{equation*}
where $p_{ij}$ defines the probability of an edge being observed between nodes $i$ and $j$. We choose to use this model, because the parameters can take any real values, whilst the the edge probabilities $p_{ij}$ are guaranteed to lie between 0 and 1. This model only deviates from the equivalent log model when the parameter values become very large - it is this effect that prevents $p_{ij}$ from reaching (and exceeding) 1. The node-specific parameters $\alpha_i$, $i\in{1,...,m}$ are elements of the parameter vector $\boldsymbol{\alpha}$ which defines a power-law degree-distribution for the nodes. Each $\alpha_i$ is generated as the logarithm of a sample taken from a bounded Pareto distribution \citep{olhede2012degree}. We note that because our $\alpha_i$ are chosen to be random, our generated networks are exchangeable \citep{kallenberg2005probabilistic}, whereas if the elements of $\boldsymbol{\alpha}$ were defined deterministically, these networks would instead be generated under the inhomogenous random graph model \citep{bollobas2007phase}. The community parameter $\theta_{ij}$ is allowed to take two values: $\theta_{ij}=\theta_{\text{in}}$ if $i$ and $j$ are in the same community, and $\theta_{ij}=\theta_{\text{out}}$ otherwise. We choose to constrain $\theta_{ij}$ in this way because it is a simple means of adding community structure, and it is equivalent to a modelling constraint which improves parameter identifiability in some formulations of the stochastic blockmodel \citep{newman2013spectral}. After generating the $p_{ij}$, the network is generated by sampling each $A_{ij}$ according to the law of:
\begin{equation*}
A_{ij}\sim\text{Bernouilli}\left(p_{ij}\right).
\end{equation*}
The communities themselves are planted in the network as randomly chosen groups of 150 nodes. We set the number of communities $k=20$, and hence the generated networks each comprise $m=3000$ nodes. 

Having generated a network with known ground-truth community structure in this way, we use it to randomly generate a sample correlation matrix $\hat{\mathbf{r}}$, from which we attempt to reproduce the known community structure. To do this, we first generate a random sample covariance matrix $\hat{\mathbf{S}}_{ij}$ for each pair of nodes $i$ and $j$, according to:
\begin{equation*}
\hat{\mathbf{S}}_{ij}\sim\text{Wishart}\left(\mathbf{S},\nu\right)
\end{equation*}
where 
\begin{equation*}
\mathbf{S}=\left( \begin{array}{cc}
1 & r_\text{gen} \\
r_\text{gen} & 1
\end{array} \right)
\end{equation*}
if $A_{ij}=1$, where $r_\text{gen}$ is the model generative correlation coefficient, and 
\begin{equation*}
\mathbf{S}=\left( \begin{array}{cc}
1 & 0 \\
0 & 1
\end{array} \right)
\end{equation*} 
if $A_{ij}=0$, and $\nu$ is the degrees of freedom. We then calculate the estimate of the sample Pearson correlation coefficient $\hat{r}_{ij}$ for nodes $i$ and $j$ as $\hat{r}_{ij}=\left(\hat{\mathbf{S}}_{ij}\right)_{12}/\sqrt{\left(\hat{\mathbf{S}}_{ij}\right)_{11}\times\left(\hat{\mathbf{S}}_{ij}\right)_{22}}=\left(\hat{\mathbf{S}}_{ij}\right)_{21}/\sqrt{\left(\hat{\mathbf{S}}_{ij}\right)_{11}\times\left(\hat{\mathbf{S}}_{ij}\right)_{22}}$. With all elements of $\hat{\mathbf{r}}$ generated in this way, with $\hat{r}_{ij}=\hat{r}_{ji}$ and $\hat{r}_{ii}=0$ for $i, j\in\left\{1,...,m\right\}$, we proceed with network inference and community detection according to the methodology set out in section \ref{adjMatEst}. 

We test the proposed methodology on networks generated with values of $\theta_{\text{in}}\in\left\{50,30,20,10\right\}$, which correspond to within-community edge density $\rho_\text{in}\in\left\{0.81,0.34,0.15,0.039\right\}$. For all networks, we set $\theta_{\text{out}}=1$, corresponding to between-community edge density $\rho_\text{out}=0.0013$. We generate sample covariance matrices with $r_\text{gen}\in(0,1]$, and degrees of freedom $\nu\in\left\{50, 100, 200\right\}$. For each combination of parameters, we carry out 50 repetitions of network generation followed by network inference and community detection. These repetitions enable assessment of the variability of the accuracy of the network inference. To compare detected communities in the inferred network with the ground-truth planted communities, we use the normalised mutual information (NMI) \citep{danon2005comparing}. The NMI assesses the numbers of nodes which appear together in the detected communities, compared with whether they appeared together in the planted communities (adjusted for group sizes). The NMI takes the value 1 if the communities are perfectly reproduced in the community detection, and 0 if they are not reproduced at all, and somewhere in between if they are partially reproduced.

\begin{figure}[ht!]
\centering
\vspace{-2ex}
\hspace{0ex}\includegraphics[width=0.8\textwidth]{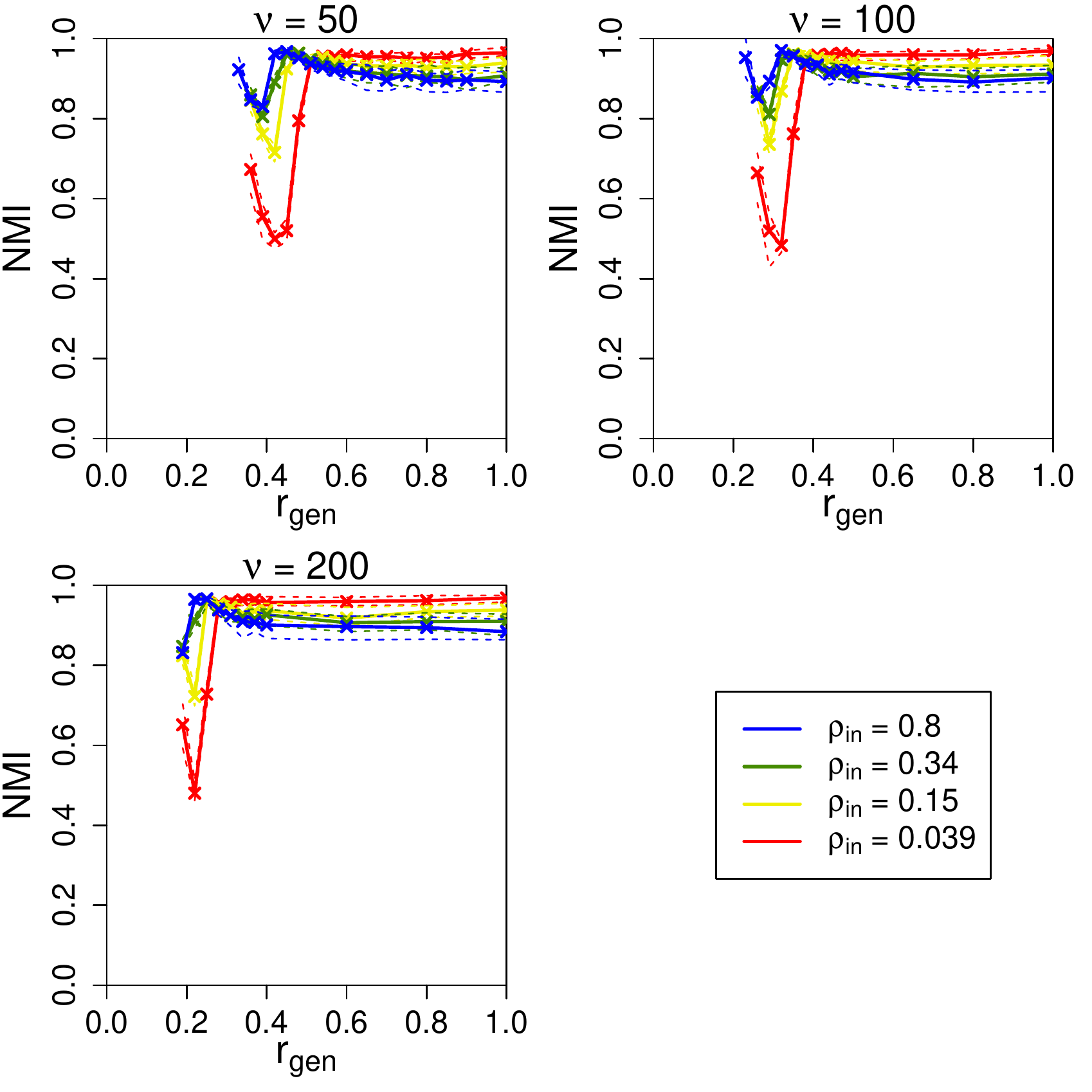}
\vspace{-2ex}
\caption{Simulation study: performance of proposed methodology.} \label{simRes}
\caption*{Normalised mutual information (NMI) compares detected community structure with ground-truth planted communities. Each line corresponds to a different within-community edge-density; these are set as $\rho_\text{in}\in\left\{0.81,0.34,0.15,0.039\right\}$ by setting $\theta_{\text{in}}\in\left\{50,30,20,10\right\}$. The degrees of freedom, $\nu$, are set as $\nu\in\left\{200,100,50\right\}$. For each network, the number of nodes $m=3000$, the ground-truth number of communities is $k=20$, and the between-community edge density is set as $\rho_\text{out}=0.0013$ by setting $\theta_{\text{out}}=1$. Dashed lines indicated quartiles.}
\vspace{3ex}
\end{figure}

The results of the simulation study are shown in Figure \ref{simRes}. The accuracy of reproduction of the ground-truth community structure is high (as evidenced by NMI values close to 1), if the generative correlation coefficient $r_\text{gen}$ is sufficiently large. There is rapid deterioration of performance below the optimal range of $r_\text{gen}$, and when $r_\text{gen}$ is sufficiently low, no edges are detected. In this regime, the non-zero mean component of the generative mixture model is centred sufficiently close to zero that the ${z}_{ij}$ from this component become categorised together with those from the zero-mean mixture component. The result is that the model fitting effectively assigns all ${z}_{ij}$ to the zero-mean component. However, as long as the generative correlation coefficient $r_\text{gen}$ is sufficiently large, the method performs well even with fairly sparse within-community edge density in the ground-truth planted communities. Typically, the method fails when $r_\text{gen}$ falls below roughly 0.45, 0.35 and 0.25 for $\nu=50$, $\nu=100$ and $\nu=200$, respectively. In the regime where the method is close to failing, there is an apparent increase in performance before complete failure, which manifests as the spikes in NMI values seen in in Figure \ref{simRes} in the range $0.3<\rho_\text{gen}<0.4$. This phenomenon occurs because in this regime, there is a transition from mainly larger communities being detected to many more smaller communities being detected, as evidenced by a decrease in the mode of the distribution of detected community sizes (Supplementary Figure S1). Community size is initially maintained in this regime as $\rho_\text{gen}$ is decreased below 0.4, and the corresponding decline in performance occurs because these larger communities only partially overlap with the ground-truth communities. As $\rho_\text{gen}$ is decreased further and gets close to the point where the methodology will fail completely, fewer edges are detected overall leading to the larger communities breaking up into many small communities. These small communities are mostly subsets of the the ground-truth communities, and this is reflected in the higher NMI values. As $\rho_\text{gen}$ is decreased beyond this regime, no edges are detected and the method fails completely. We also note that for large values of $r_\text{gen}$, the performance of the methodology is slightly worse for the largest values of $\rho_\text{in}$. The planted ground-truth communities each comprise 150 nodes, and this decrease in performance occurs because in this regime several of these communities coalesce in the inferred network to form a much larger connected component (Supplementary Figure S2). This is likely to be due to the higher false-positive rate in this regime (Supplementary Figure S4) leading to spurious connections between communities.

The thresholding methodology which underlies the proposed methodology of section \ref{modelFitSect} was originally developed in the context of thresholding data vectors \citep{johnstone2004needles}. Applying this methodology to relational data matrices such as covariance and correlation matrices is complicated by the presence of additional dependency structure, and to mitigate spurious detection, the conservative adjacency matrix estimate of equation \ref{eBayesMixMLE} is used. To check the performance of the methodology in this context of adjacency matrix thresholding against the intended vector thresholding application, we carried out comparative true positive rate (sensitivity) and false positive rate (1-specificity) analyses. For these analyses the same simulated data is considered as is presented in Figure \ref{simRes}, and the results appear in the supplement in Figures S3 and S4. True and false positive rates are calculated for the adjacency matrix inference presented in sections \ref{covCorSubSect} - \ref{modelFitSect}, and these results are labelled `matrix' in Figures S3 and S4. The equivalent results based on equation \ref{eBayesMixMLEvec} are also recorded for each row of the thresholded adjacency matrix before applying the conservative estimate of equation \ref{eBayesMixMLE}, and the means of these over each row of the adjacency matrix are also shown in Figures S3 and S4 and labelled `vector'. The true positive rate is only slightly lower for adjacency matrix inference than for vector thresholding, except when $\rho_\text{in}$ is lowest. The false positive rate is close to zero in all cases, although it is apparently sufficiently great for the largest values of $\theta_{\text{in}}$ and $\rho_\text{in}$ to cause spurious coalescence of some communities, as discussed.

\subsection{Comparison with popular clustering methods}
The clustering problem is fundamentally different to that of community detection, although there are nevertheless many similarities. The basic task of clustering is to group together entities (usually variables or samples) based on their similarity or distance from one another in observation space, which can assessed by, for example, Pearson correlation. When the entities being grouped are nodes in a network, the problems of clustering and community detection are very similar. In this study, we infer binary-valued networks from continuous data before carrying out community detection. However, a number of popular methods provide alternative means of clustering entities into groups (which may be considered equivalent to communities) based on continuous data.

\begin{figure}[ht!]
\centering
\vspace{-2ex}
\includegraphics[width=0.8\textwidth]{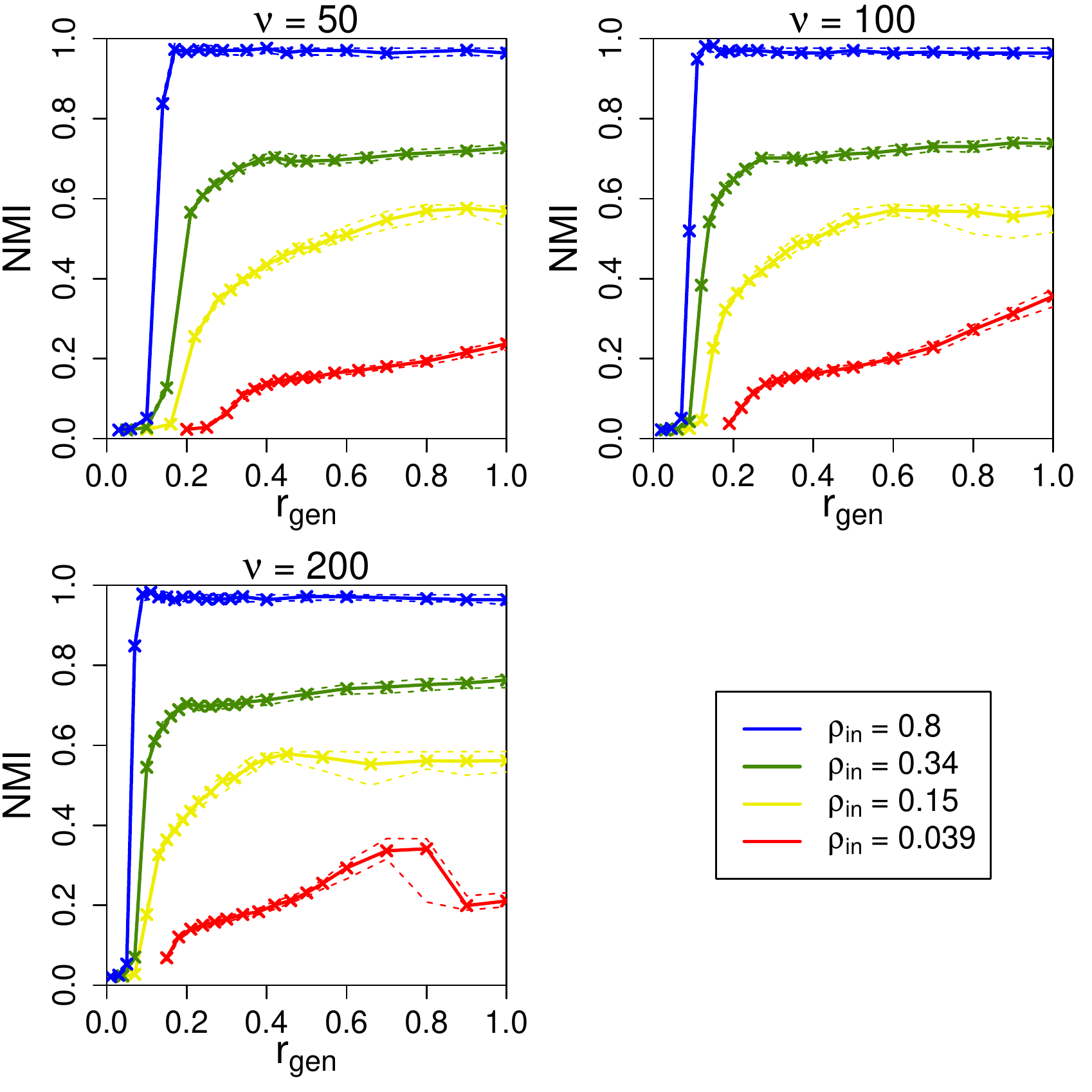}
\vspace{-2ex}
\caption{Simulation study: spectral clustering without network inference.} \label{simResSpecClust}
\caption*{Normalised mutual information (NMI) compares detected community structure with ground-truth planted communities. Each line corresponds to a different within-community edge-density; these are set as $\rho_\text{in}\in\left\{0.81,0.34,0.15,0.039\right\}$ by setting $\theta_{\text{in}}\in\left\{50,30,20,10\right\}$. The degrees of freedom, $\nu$, are set as $\nu\in\left\{200,100,50\right\}$. For each network, the number of nodes $m=3000$, the ground-truth number of communities is $k=20$, and the between-community edge density is set as $\rho_\text{out}=0.0013$ by setting $\theta_{\text{out}}=1$. Dashed lines indicated quartiles.}
\vspace{3ex}
\end{figure}

A method of clustering which is very popular across the biological and social sciences is hierarchical clustering. In that method, variables or samples are grouped together according to their distance from one another. A popular measure of distance between a pair $i$ and $j$ of such variables or samples is $1-|\hat{r}_{ij}|$, where $|\hat{r}_{ij}|$ is the absolute value of the Pearson correlation coefficient between $i$ and $j$ estimated from the available observations. Hence, this method can be easily applied to data of the type presented here (without carrying out the network inference presented in section \ref{modelFitSect}). We tested this method on the simulated data presented in section \ref{simStudSect}, by applying hierarchical clustering to the generated sample correlation matrix $\hat{\mathbf{r}}$ before comparing the detected clusters with the planted communities. However, we found that in every case, the result of this comparison was an NMI value close to 0. Therefore, we may conclude that hierarchical clustering performs significantly worse than the methods presented here on problems of this type.

One of the most popular clustering methods is $K$-means. In that method samples (which may be thought of as equivalent to network nodes) are grouped into $K$ clusters based on their location in $N$-dimensional space. On its own, this method is fundamentally ill-suited to network data because of the high dimensionality of the problem. However, $K$-means clustering is often used in spectral clustering after dimension reduction by SVD: we use that method of spectral clustering in this paper to fit the stochastic blockmodel. Spectral clustering can also be used to cluster continuous data, and so for comparison we have applied regular spectral clustering (without carrying out the network inference described in sections \ref{covCorSubSect} - \ref{modelFitSect}) to the simulated data presented in section \ref{simStudSect}. To do this, we applied spectral clustering as described at the start of section \ref{examples} directly to $|\hat{\mathbf{r}}|$, the absolute of the generated sample correlation matrix (i.e. to continuous data). The absolute values are used to ensure that the data is non-negative, as required for spectral clustering \citep{von2007tutorial}. The results appear in Figure \ref{simResSpecClust}. Spectral clustering applied directly to $\hat{\mathbf{r}}$ is generally less accurate (according to the NMI) than if the network inference/thresholding of sections \ref{covCorSubSect} - \ref{modelFitSect} is first applied (Figure \ref{simRes}). One exception when spectral clustering applied directly to $\hat{\mathbf{r}}$ is more accurate occurs when $r_\text{gen}$ is lowest, as in that regime the problem of total failure of the network inference/thresholding (as discussed in section \ref{simStudSect}) is avoided. Another such exception occurs when $\rho_\text{in}$ is highest and $r_\text{gen}$ is large. The reason is that in this regime, the phenomenon of the ground-truth clusters/communities coalescing due to false positives caused by the network inference/thresholding (also as discussed in section \ref{simStudSect}) is avoided. However in general, for problems of the type presented here, applying the network inference/thresholding of sections \ref{covCorSubSect} - \ref{modelFitSect} prior to carrying out spectral clustering produces more accurate results. Furthermore, as this network inference/thresholding generally results in a sparse adjacency matrix, it allows use of efficient computational methods to find the top components in the SVD which are required for spectral clustering.

\subsection{Genomics example}
We now give an illustrative example of a practical application of these methods to a standard problem in genomics. Community detection can be used to infer groups of genes which comprise functional subnetwork modules, or groups of co-regulated genes. Examples of such groups are found in gene regulatory networks and protein signalling networks \citep{shen2002network}. Defining $\mathbf{x}(k)$ to be the gene expression measurements for sample $k$ for the genes $x_1,x_2,...,x_m$, we calculate the covariance matrix according to equation \ref{covMat}, and carry out network inference as described in sections \ref{covCorSubSect} - \ref{modelFitSect}. We note that the network edges detected in this way may be transitive edges, i.e. they do not necessarily represent physical interactions between genes and gene products. To determine this would require additional functional data, such as those relating to DNA binding by gene products (e.g., transcription factors) \citep{jojic2013identification}. However, in general the groups of genes detected in this way can be expected to form biologically meaningful subnetwork modules, generating biological hypotheses which may justify further investigation by experimental scientists. 

\begin{figure}[ht!]
\centering
\vspace{-2ex}
\includegraphics[width=0.85\textwidth]{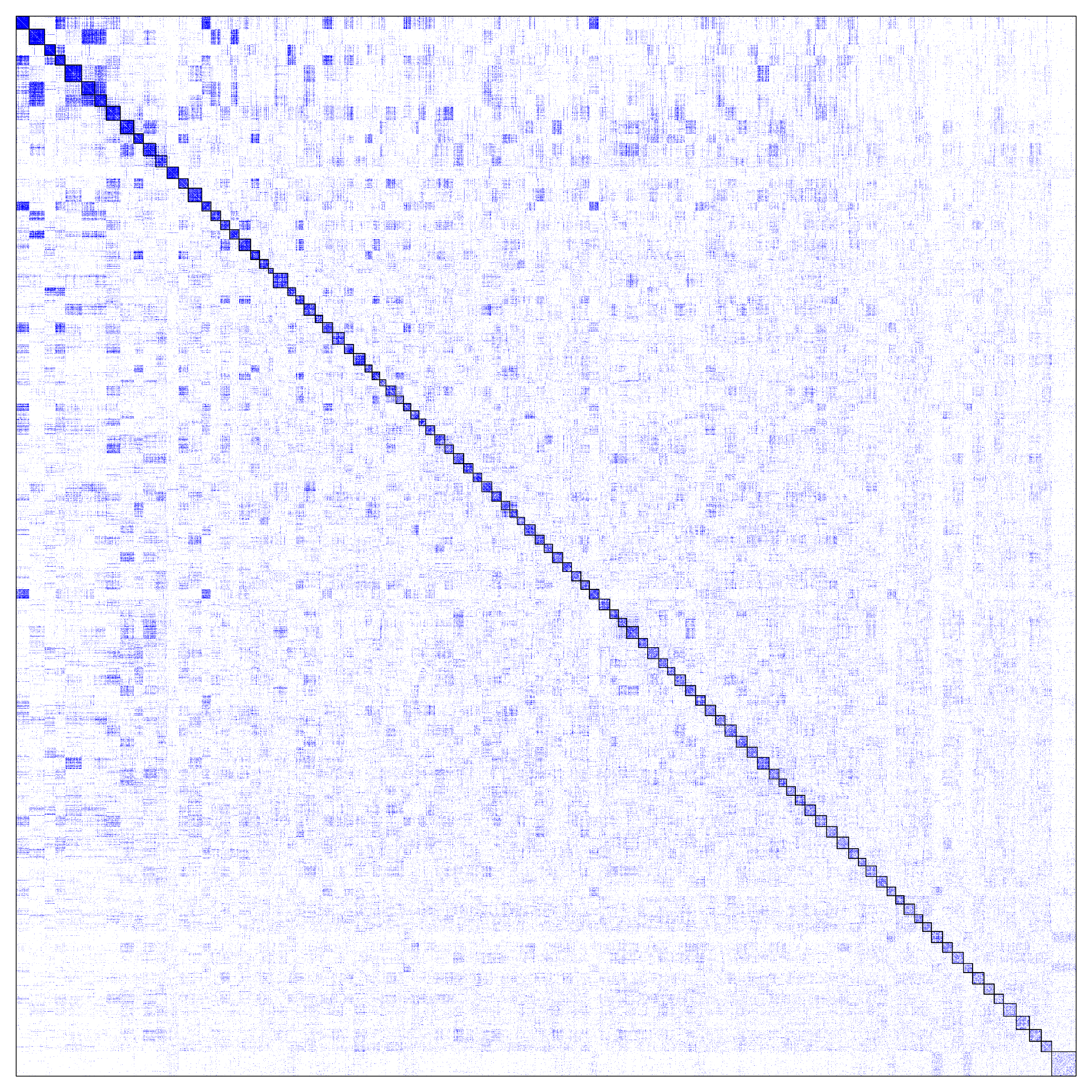}
\vspace{-2ex}
\caption{Detected communities in a gene expression data set, relating to lung cancer.} \label{geneExprAdj}
\caption*{Entries in the adjacency matrix equal to 1 (representing a network edge) are coloured blue, and detected communities are outlined in black.}
\vspace{3ex}
\end{figure}

We carried out this process of network inference and community detection in gene expression data from 8 different types of cancer: brain, breast, colon, kidney, lung, ovarian, rectal and uterine (data source: The Cancer Genome Atlas \citep{hampton2006cancer}). Each data set comprises gene expression measurements for 17505 genes (i.e., $m=17505$). Figure \ref{geneExprAdj} shows the inferred adjacency matrix after community detection for the lung cancer data-set. The number of communities is estimated as 105 by the network histogram method \citep{olhede2014net} for this data-set, and the edge density is $\rho=0.062$ (which is typical of all 8 gene expression datasets).

We also tested the domain-relevance of the communities detected in the inferred networks. We tested the overlap of the genes of each detected community separately with each of $10295$ known gene-groups (data source: \allowbreak\textit{http://www.broadinstitute.org/gsea/msigdb/}). This is known as `gene set enrichment analysis' (GSEA) \citep{subramanian2005gene}. Table \ref{comEnrichTab} shows the percentage of the communities detected in each cancer data-set which overlapped significantly with at least one of these known gene-groups. For this purpose, significance is assessed by Fisher's exact test, with the significance level set by FDR (false discovery rate) adjusted $p<0.05$.
\begin{table}[ht!]
\begin{center}
{\small
\begin{tabular}{|c|c|c|c|c|c|c|c|}
\hline
Breast & Colon & Brain & Kidney & Lung & Ovarian & Renal & Uterine \\
\hline
97\% & 86\% & 87\% & 76\% & 89\% & 96\% & 76\% & 66\% \\
\hline
\end{tabular}
}
\caption{Domain-relevance of detected communities in the genomics example.} 
\caption*{The table shows the percentage of the communities detected in each cancer data-set which overlap significantly (Fisher's exact test, FDR-adjusted $p<0.05$) with at least one known gene group.}\label{comEnrichTab}
\end{center}
\vspace{-2ex}
\end{table}
As a benchmark, we also sampled random groups of genes from the $17505$ genes represented in the cancer data-sets, and tested them for overlap with the same $10295$ known gene-groups. The number of genes in each random sample was itself randomly sampled from the distribution of the sizes of the communities detected in the cancer data-sets. We took 1000 randomly sampled groups of genes like this, of which $2\%$ overlapped significantly with at least one of the known gene-groups. These results show a high level of domain-relevance of the detected communities, in all 8 genomics data-sets analysed here.

\subsection{Consumer product review example}
We now give a second, contrasting illustrative example of a practical application of these methods to real data, based on a consumer-product review dataset. We downloaded movie review data from the \textit{Movie Lens} database, which details $1\ 000\ 209$ reviews of 3952 different movies, by 6040 unique users who each provided at least 20 different reviews (data source: \allowbreak\textit{http://grouplens.org/datasets/movielens/}). Covariate information is also available, classifying each user into one of 7 age groups and 20 professions; this can be used to verify the detected communities/clusters.

For each pair of users $(i,j)$, we tested the overlap of the movies reviewed by user $i$ with the movies reviewed by user $j$ with Fisher's exact test. This provided an estimated \textit{p}-value for each pair of users ${p}_{ij}$, under the null hypothesis that there is no significant overlap between the movies reviewed by users $i$ and $j$. These are a one-tailed test \textit{p}-values corresponding to an alternative hypothesis that there is more overlap between movies reviewed by users $i$ and $j$ than would be expected by chance. Then, we applied the inverse normal transformation to each ${p}_{ij}$ to obtain the values of ${z}_{ij}$, and obtained the estimate of the adjacency matrix $\hat{\mathbf{A}}$ as described in sections \ref{covCorSubSect} - \ref{modelFitSect}. Using the network histogram method \citep{olhede2014net}, the optimal number of communities for the blockmodel was estimated as 125. However the granularity of this estimate is much greater than that of the covariate information we have available for verification of detected clusters. The network histogram method estimates the optimal granularity for the stochastic blockmodel, however we can also select a smaller number of communities with which to fit the stochastic blockmodel, whilst noting that this will not result in the optimal blockmodel as assessed by the mean squared integrated error (MISE) \citep{olhede2014net}.  We selected 15 communities for the blockmodel, which is of the same order as the number of covariate classes, but chosen to be less than the total number of classes to take account of the fact that many of these classes are overlapping. The edge density $\rho$ for the inferred adjacency matrix $\hat{\mathbf{A}}$ is calculated as $\rho=0.16$, which is relatively high.

\begin{figure}[ht!]
\centering
\vspace{-2ex}
\includegraphics[width=0.99\textwidth]{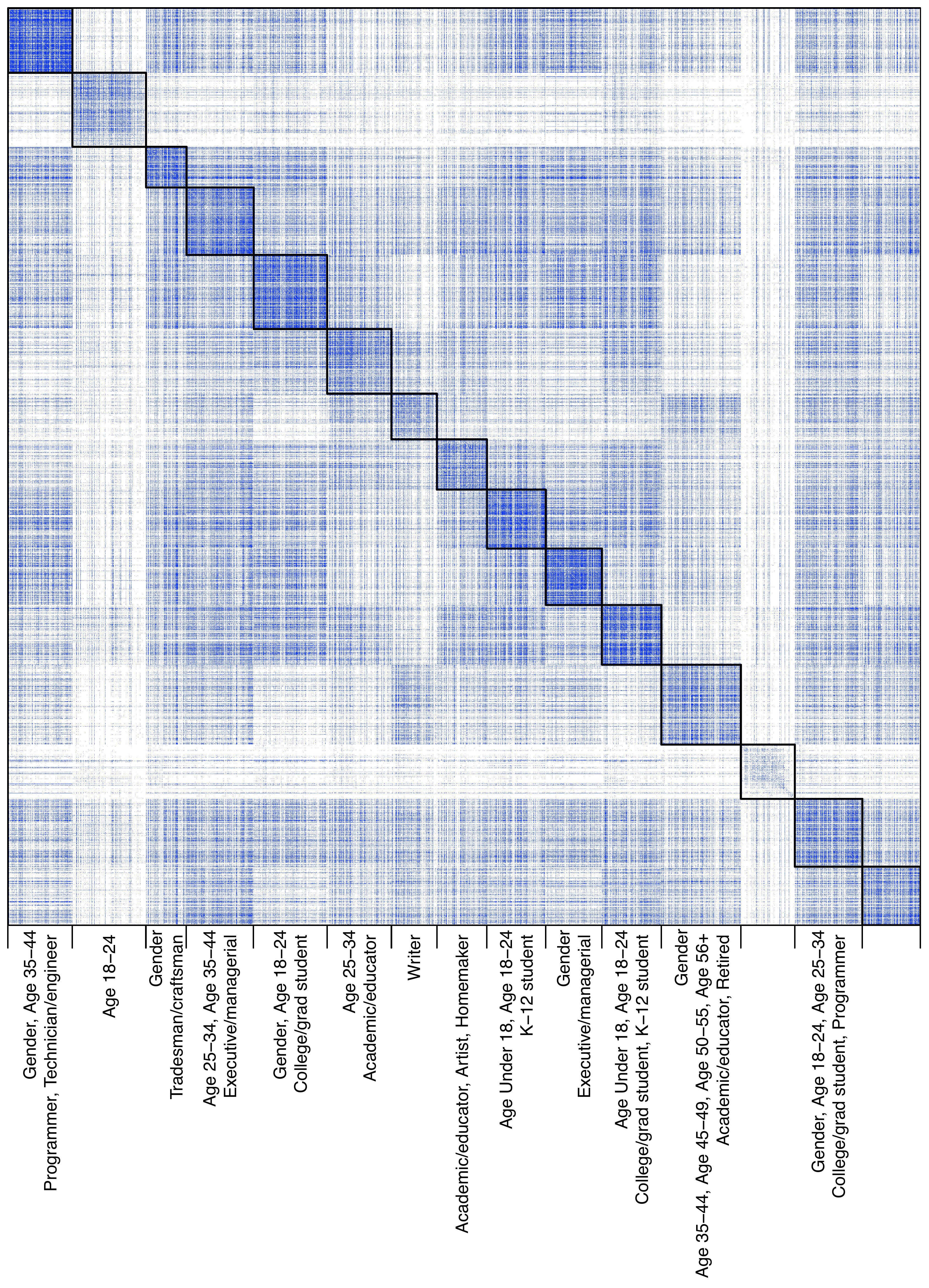}
\vspace{-2ex}
\caption{Detected communities in the movie review data set.} \label{movieLensAdj}
\caption*{Entries in the adjacency matrix equal to 1 (representing a network edge) are coloured blue, and detected communities are outlined in black.}
\vspace{3ex}
\end{figure}

Figure \ref{movieLensAdj} shows the inferred adjacency matrix after community detection. The detected communities are tested for overlap with the known covariate groups; those which overlap significantly (Fisher's exact test, FDR-corrected $p<0.05$) are specified along the margin. Almost all of the detected communities/clusters overlap with at least one covariate group, and several communities/clusters overlap with multiple covariate groups. Where the overlap is with multiple covariate groups, there is generally an obvious link between these groups, such as similar age groups, or professions which suggest similar demographic groups. These findings show that this methodology is very effective in the context of this example, in which we obtain $\hat{\mathbf{A}}$ from an arbitrary non-Gaussian distribution, based on corresponding \textit{p}-values of association ${p}_{ij}$ between pairs of variables $(x_i,x_j)$. 

\section{Conclusion}
In this paper, we have proposed methodology combining estimation of binary-valued adjacency matrices with community detection via the stochastic blockmodel, based on sample covariance and correlation matrices or more general test statistics quantifying association between pairs of variables. We have presented the theoretical basis for this proposed methodology, and provided practical details for its implementation. We have shown the accuracy of this methodology in the context of a simulation study, and have shown its effectiveness in several contexts based on multiple real data-sets, with a range of sparsities. We have also shown that this methodology performs better than popular clustering methods for discovering latent groupings in data of the type presented here. An important point to note, is that some network edges inferred from the correlation structure of data as in the methodology proposed here may be what are often referred to as `transitive edges'. I.e., an inferred edge may not correspond to a direct physical real-life interaction, instead deriving from some indirect interaction which may alternatively be mediated via a less direct route through the network, possibly also involving unobserved variables. Interesting extensions to this methodology include consideration of overlapping blocks in the stochastic blockmodel \citep{latouche2011overlapping}, and development of an online version of the methodology as a computationally efficient approach to large and growing data-sets \citep{zanghi2010strategies}. This methodology would be expected to work equally well in many other networks contexts, and in more general scenarios where the aim is to cluster together correlated variables. This methodology can be implemented using readily available and computationally efficient algorithms, and performs well on large high-dimensional datasets.
\clearpage
\bibliography{../references}

\section*{Acknowledgements}
TB acknowledges support from the UK EPSRC and MRC via UCL CoMPLEX.

\end{document}